\documentclass[12pt]{article}
\usepackage{latexsym,amsmath,righttag,chicago,psfig,doublespace,numinsec,graphicx}

\title{\bf Penalized Clustering of Large Scale Functional Data with Multiple Covariates}
\author{{\sc Ping Ma and Wenxuan Zhong }\thanks{ Ping Ma and Wenxuan Zhong are Assistant Professors (E-mail:
pingma, wenxuan@uiuc.edu), Department of Statistics, University of
Illinois, Champaign, IL 61820. PM's research was partially supported
by National Science Foundation grant DMS-0723759. The authors are
grateful to Jun S. Liu, Chong Gu, Yu Zhu, Xuming He, Steve Portnoy
for many illuminating discussions on this article. The authors thank
Kurt Kwast for providing the yeast microarray data. The authors also
thank the editor, the associate editor, the two referees, John
Marden, and Adam Martinsek for their constructive comments and
suggestions that have led to significant improvement of this
article. }}
\date{}
\newcommand{\mbf}{\mathbf}
\newcommand{\sbf}{\boldsymbol}

\begin{document}
\maketitle

\begin{abstract}

In this article, we propose a penalized clustering method for large
scale data with multiple covariates through a functional data
approach.  In the proposed method, responses and covariates are
linked together through nonparametric multivariate functions (fixed
effects), which have great flexibility in modeling a variety of
function features, such as jump points, branching, and periodicity.
Functional ANOVA is employed to further decompose multivariate
functions in a reproducing kernel Hilbert space and provide
associated notions of main effect and interaction. Parsimonious
random effects are used to capture various correlation structures.
The mixed-effect models are nested under a general mixture model, in
which the heterogeneity of functional data is characterized. We
propose a penalized Henderson's likelihood approach for
model-fitting and design a rejection-controlled EM algorithm for the
estimation. Our method selects smoothing parameters through
generalized cross-validation. Furthermore, the Bayesian confidence
intervals are used to measure the clustering uncertainty. Simulation
studies and real-data examples are presented to investigate the
empirical performance of the proposed method. Open-source code is
available in the R package MFDA.


{\it Running title:} Penalized clustering of functional data.
\end{abstract}

\section{Introduction}

With the rapid advancement in high throughput technology, extensive
repeated measurements have been taken to monitor the system-wide
dynamics in many scientific investigations. A typical example is
temporal gene expression studies, in which a series of micorarray
experiments are conducted sequentially during a biological process,
e.g., cell cycle microarray experiments
(\shortciteNP{spellman:1998}). At each time point, mRNA expression
levels of thousands of genes are measured simultaneously. Collected
over time, a gene's ``temporal expression profile'' gives the
scientist some clues on what role this gene might play during the
process. A group of genes with similar profiles are often
``co-regulated'' or participants of a common and important
biological function. Thus clustering genes into homogeneous groups
is a crucial first step to decipher the underlying mechanism. The
need to account for intrinsic temporal dependency of repeated
observations within the same individual renders traditional methods
such as K-means and hierarchical clustering inadequate. By casting
repeated observations as multivariate data with certain correlation
structure, one ignores the time interval and time order of sampling.
Additionally, missing observations in the measurements yield an
unbalanced design, which requires imputation beforehand for
application of multivariate approaches, e.g., the multivariate
Gaussian clustering method (MCLUST, \citeNP{fraley:2002}).

In addition to the time factor, such repeated measurements often
contain other covariates, e.g., replicates at each time point,
species in comparative genomics studies (\shortciteNP{haoli:04}),
and treatment groups in case-control studies
(\shortciteNP{storey:05}), as well as many factors in a factorial
designed experiment. Incorporation of multiple covariates adds
another layer of complexity. Clustering methods taking all these
factors into account are still lacking.

Recently, nonparametric analysis of data in the form of curves, i.e.
functional data, is subject to active research. See
\citeANP{Rams:Silv:func:2005} (\citeyearNP{Rams:Silv:func:2005},
\citeyearNP{Rams:Silv:appl:2002}) for a comprehensive treatment of
functional data analysis. A curve-based clustering method (FCM) was
introduced in \citeN{Jame:Suga:2003} to cluster sparsely sampled
functional data. Similar approaches were developed in
\shortciteANP{luan:03} (\citeyearNP{luan:03}, \citeyearNP{luan:04})
and \shortciteN{heard:06} to analyze temporal gene expression data.
 Although these methods model the time factor explicitly, none
of them  are designed to accommodate additional factors. Moreover,
smoothing-related parameters, e.g., knots and degrees of freedom, in
these methods  are the same across all clusters and must be
specified {\it a priori}. Consequently, they can not model
drastically different patterns among different clusters, which leads
to high false classification rate. Finally, the computational costs
of these methods are very high for large scale data.

Motivated by analysis of temporal gene expression data, we propose a
flexible functional data clustering method that overcomes the
aforementioned obstacles. In our proposed method, responses and
covariates are linked together through nonparametric multivariate
functions (fixed effects), which have great flexibility in modeling
a variety of function features, such as jump points, branching, and
periodicity. Functional ANOVA is employed to further decompose
multivariate functions (fixed effects) in a reproducing kernel
Hilbert space and provide associated notions of main effect and
interaction (\citeNP{wahba:90} and \citeNP{gu:02}). Parsimonious
random effects, complementing the fixed effects, are used to capture
various correlation structures. The mixed-effect models are nested
under a general mixture model, in which the heterogeneity of the
functional data is characterized. We propose a penalized Henderson's
likelihood approach for model-fitting and design a
rejection-controlled EM algorithm for estimation. In this EM
algorithm, the E-step is followed by a rejection-controlled sampling
step (\shortciteNP{Liu:Chen:Wong:1998}) to eliminate a significant
number of functional observations, whose posterior probabilities of
belonging to a particular cluster is negligible, from calculation in
the subsequent M-step.  The M-step is decomposed into the
simultaneous maximization of penalized weighted least squares in
each cluster. The smoothing parameters associated with the penalty
are selected by generalized cross-validation, which can be shown to
track a squared error loss asymptotically. Our method is thus
data-adaptive and automatically captures some important functional
fluctuations. For model selection, we employ BIC to select the
number of clusters. Moreover, the proposed method not only provides
subject-to-cluster assignment but also the estimated mean function
and associated Bayesian confidence intervals for each cluster.  The
Bayesian confidence intervals are used to measure the clustering
uncertainty. These nice features make the proposed method extremely
powerful for clustering large scale functional data.

The remainder of the article is organized as follows. In Section 2,
we present a nonparametric mixed-effect model representation for
functional data. A mixture model for clustering is considered in
Section 3. Simulation and real data analysis follow in Section 4 and
5. A few remarks in Section 6 conclude the article. Proofs of the
theorems are collected in Appendix.

\section{Nonparametric Mixed-Effect Representation of Homogeneous Functional Data }

Assuming the data are homogeneous, i.e., the number of clusters is
one, we shall present a mixed-effect representation of functional
observations.

\subsection{The Model Specification}

We assume the functional data of the $i$th individual,
$\sbf{y}_i=(y_{i1}, \cdots, y_{in_i})^{T}$, follows the mixed-effect
model,
\begin{equation}\label{mixedn}
\sbf{y}_i= \mu(\mbf{x}_i)+ \mbf{Z}_i\mbf{b}_i+ \sbf{\epsilon}_i,
\end{equation}
where the population mean $\mu$ is assumed to be a smooth function
defined on a generic domain $\Gamma$,
$\mbf{x}_i=(x_{i1},\ldots,x_{in_i})^{T}$ is an ordered set of
sampling points, $\mbf{b}_i\sim{N}(0,\mbf{B})$ is a $p\times{1}$
random effect vector associated with a $n_i\times p$ design matrix
$\mbf{Z}_i$, and random errors
$\sbf{\epsilon}_i\sim{N}(0,\sigma^{2}\mbf{I})$ are independent of
$\mbf{b}_i$'s, and of each other. The random effect covariance
matrix $\mbf{B}$ and random error variance $\sigma^{2}$ are to be
estimated from the data. Model (\ref{mixedn}) has been extensively
studied in the statistical literature. See \citeN{wang:98a},
\shortciteN{Zhan:Lin:Raz:Sowe:semi:1998}, \citeN{gm:02},  and
references therein.

For multivariate $x$ where  $x=(x_{\langle 1 \rangle}, x_{\langle 2
\rangle}, \cdots, x_{\langle d \rangle})^{T}$,
 each entry $x_{\langle k \rangle}$
takes values in some fairly general domain ${\Gamma_k}$, i.e.,
${\Gamma}=\otimes_{k=1}^{d}\Gamma_k $. Some examples are
\begin{ex}\rm
$\Gamma=[0, {\cal{T}}]\times \{ 1,\cdots, c \}$ to model temporal
variation from  time $0$ to time ${\cal{T}}$  under multiple
conditions; $\Gamma=\text{Circle} \times \{ 1,\cdots, s \}$ to model
periodicity of a biological process of multiple species.
\end{ex}

The functional ANOVA decomposition of a multivariate function $\mu$
is
\begin{equation}\label{fanoav}
    \mu(x)=\mu_0+\sum_{j=1}^{d}\mu_j(x_{\langle j \rangle})+\sum_{j=1}^{d}\sum_{k=j+1}^{d}\mu_{jk}(x_{\langle j \rangle},
    x_{\langle k \rangle})+ \cdots + \mu_{1,\cdots, d}(x_{\langle 1 \rangle}, \cdots,x_{\langle d \rangle})
\end{equation}
where $\mu_0$ is a constant, $\mu_j$'s are the main effects,
$\mu_{jk}$'s are the two-way interactions, and so on. The
identifiability of the terms  in (\ref{fanoav}) is assured by side
conditions through averaging operators. See \citeN{wahba:90} and
\citeN{gu:02}.

By using different specifications of the random effect $\mbf{b}_i$
and associated design matrix $\mbf{Z}_i$, model (\ref{mixedn}) can
accommodate various correlation structures.
\begin{ex}\rm
If we let $p=1$,  i.e.,  $\mbf{b}_i$ is a scalar,
$\mbf{B}=\sigma_{b}^2$ and $\mbf{Z}_i=\mbf{1}$, we have the same
correlation across time. If we let $p=2$, i.e.
$\mbf{b}_i=(b_{i1},b_{i2})^{T}$, $\mbf{B}=
\begin{pmatrix}
\sigma^2_{b_1}     & \sigma^2_{b_1b_2} \\
\sigma^2_{b_1b_2}  & \sigma^2_{b_2}
\end{pmatrix}
$ and $\mbf{Z}_i=(\mbf{1}, \mbf{x}_i)$, the difference between the
$i$th subject profile and the mean profile is a linear function in
time. The covariance between expression values at $x_1$ and $x_2$
for the same individual is
$\sigma^2_{b_1}+(x_1+x_2)\sigma^2_{b_1b_2}+x_1x_2\sigma^2_{b_2}$.
\end{ex}

\subsection{Estimation}

Model (\ref{mixedn}) is estimated using penalized least squares
through the minimization of
\begin{equation}\label{pe_hls}
\sum_{i=1}^{n}(\sbf{y}_{i}-\mu(\mbf{x}_i)-\mbf{Z}_i\mbf{b}_i)^{T}(\sbf{y}_{i}-\mu(\mbf{x}_i)-\mbf{Z}_i\mbf{b}_i)
+\sum_{i=1}^{n}\sigma^{2}\mbf{b}_{i}^{T}{\mbf{B}}^{-1}\mbf{b}_{i}+N{\lambda}
M(\mu),
\end{equation}
for $N=\sum_i{n_i}$, where the first term measures the fidelity of
the model to the data, $M(\mu)=M(\mu, \mu)$ is a quadratic
functional that quantifies the roughness of $\mu$,  and $\lambda$ is
the smoothing parameter that controls the trade-off between the
goodness-of-fit and the smoothness of $\mu$. (\ref{pe_hls}) is also
referred to as penalized Henderson's likelihood since the first two
terms are proportional to the joint density (Henderson's likelihood)
of $(\sbf{y}_{i},\mbf{b}_{i})$ (\citeNP{robin:91}).

To minimize (\ref{pe_hls}), we only need to consider smooth
functions in the space $\{ \mu: M(\mu)< \infty \}$ or subspace
therein. As a abstract generalization of the vector spaces used
extensively in multivariate analysis, Hilbert spaces inherit many
nice properties of the vector spaces. However, the Hilbert space is
too loose to use for functional data analysis since even the
evaluation functional $[x](f)=f(x)$, the simplest functional one may
encounter, is not guaranteed to be continuous in a general Hilbert
space. An example is that in the Hilbert space of square integrable
functions defined on [0,1], evaluation is not even well defined.
Consequently, one may focus on a constrained Hilbert space for which
the evaluation functional is continuous. Such a Hilbert space is
referred to as a reproducing kernel Hilbert space (RKHS), for which
\citeN{Rams:Silv:func:2005} suggested a nickname: continuous Hilbert
space. For example, the space of functions with square integrable
second derivatives is an RKHS if it is equipped with appropriate
inner products (\citeNP{gu:02}). For the evaluation functional
$[x](\cdot)$, by the Riesz representation theorem, there exists a
non-negative definite bivariate function $R(x,y)$, the reproducing
kernel, which satisfies $ \langle R(x,\cdot), f(\cdot)
\rangle=f(x)$, called the ``representer" of $[x](\cdot)$, in RKHS.
Given an RKHS, we may derive the reproducing kernel from the Green's
function associated with the quadratic functional $M(\mu)$. Since
the construction of reproducing kernel is beyond the scope of this
article, readers may refer to \citeN{wahba:90} and \citeN{gu:02} for
details.

 The minimization of (\ref{pe_hls}) is performed in a
reproducing kernel Hilbert space $\mathcal{H}\subseteq\{\mu:
M(\mu)<\infty\}$ in which $M(\mu)$ is a square semi norm. To
incorporate (\ref{fanoav}) in estimating multivariate functions, we
consider $\mu_j \in {\cal{H}}_{\langle j \rangle}$, where
${\cal{H}}_{\langle j \rangle}$ is a reproducing kernel Hilbert
space with tensor sum decomposition ${\cal{H}}_{\langle j
\rangle}={\cal{H}}_{0\langle j \rangle} \oplus {\cal{H}}_{1 \langle
j \rangle}$ where ${\cal{H}}_{0\langle j \rangle}$ is the
finite-dimensional ``parametric" subspace consisting of parametric
functions, and ${\cal{H}}_{1 \langle j \rangle}$ is  the
``nonparametric" subspace consisting of smooth functions. The
induced tensor product space is
$${\cal{H}}=\otimes_{j=1}^{d}{\cal{H}}_{\langle j
    \rangle}=\oplus_{\cal{S}}[(\otimes_{j \in \cal{S}}{\cal{H}}_{1\langle j
    \rangle})\otimes ( \otimes_{j \notin \cal{S}}{\cal{H}}_{0\langle j
    \rangle})]=\oplus_{\cal{S}}{\cal{H}}_{\cal{S}},$$
where the summation runs over all subsets ${\cal{S}} \subseteq
\{1,\cdots, d\}$. These subspaces ${\cal{H}}_{\cal{S}}$ form two
large subspaces: $\mathcal{N}_{M}=\{\eta:M(\mu)=0\}$, which is the
null space of $M(\mu)$, and $\mathcal{H}\ominus\mathcal{N}_{M}$ with
the reproducing kernel $R_{M}(\cdot,\cdot)$. The solution of
(\ref{pe_hls}) has an expression

\begin{equation}\label{ss_expr}
\mu(x)=\sum_{\nu=1}^{m}d_{\nu}\phi_{\nu}(x)
+\sum_{i=1}^{T}{c}_{i}R_{M}(s_{i},x),
\end{equation}
where $\{\phi_{\nu}\}_{\nu=1}^{m}$ is a basis of $\mathcal{N}_{M}$,
and $d_{\nu}$ and $c_i$ are the coefficients, $\mbf{s}=(s_1,\cdots,
s_T)$ is a distinct combination of all $x_{ij}(i=1, \cdots,n,
j=1,\cdots, n_i)$.

\begin{ex}\label{trt:e}\rm
Consider the temporal variation under  $a$ treatments. Take the
fixed effect as $\mu(t,\tau)$, where $\tau\in\{1,\dots,a\}$ denotes
the treatment levels.  One may decompose
\[
\mu(t,\tau)=\mu_{\emptyset}+\mu_{1}(t)
+\mu_{2}(\tau)+\mu_{1,2}(t,\tau),
\]
where $\mu_{\emptyset}$ is a constant, $\mu_{1}(t)$ is a function of
$t$ satisfying $\mu_{1}(0)=0$, $\mu_{2}(\tau)$ is a function of
$\tau$ satisfying $\sum_{\tau=1}^{a}\mu_{2}(\tau)=0$, and
$\mu_{1,2}(t,\tau)$ satisfies $\mu_{1,2}(0,\tau)=0$, $\forall\tau$,
and $\sum_{\tau=1}^{a}\mu_{1,2}(t,\tau)=0$, $\forall{t}$.  The term
$\mu_{\emptyset}+\mu_{1}(t)$ is the ``average variation'' and the
term $\mu_{2}(\tau)+\mu_{1,2}(t,\tau)$ is the ``contrast
variation''.

For flexible models, one may use
\begin{equation}\label{mixpenal}
M(\mu)=\theta_{1}^{-1}\int_{0}^{\cal{T}}(d^{2}\mu_{1}/dt^{2})^{2}dt
+\theta_{1,2}^{-1}\int_{0}^{\cal{T}}\sum_{\tau=1}^{a}(d^{2}\mu_{1,2}/dt^{2})^{2}dt,
\end{equation}
 which has a null space $\mathcal{N}_{M}$ of dimension $2a$.  A set
of $\phi_{\nu}$ are given by
\[
\{1,t,I_{\{j\}}(\tau)-1/a,(I_{\{j\}}(\tau)-1/a)t,j=1,\dots,a-1\},
\]
and the function $R_{M}$ is given by
\[
R_{M}(t_{1},\tau_{1};t_{2},\tau_{2})
=\theta_{1}\int_{0}^{\cal{T}}(t_{1}-u)_{+}(t_{2}-u)_{+}du
+\theta_{1,2}(I_{\{\tau_{1}\}}(\tau_{2})-1/a)\int_{0}^{\cal{T}}(t_{1}-u)_{+}(t_{2}-u)_{+}du
\]
See, e.g., \citeANP{gu:02} (\citeyearNP{gu:02}, $\S$2.4.4). To force
an additive model
\begin{equation}\label{addspline}
\mu(t,\tau)=\mu_{\emptyset}+\mu_{1}(t)+\mu_{2}(\tau),
\end{equation}
which yields parallel curves at different treatments, one may set
$\theta_{1,2}=0$ and remove $(I_{\{j\}}(\tau)-1/a)t$ from the list
of $\phi_{\nu}$.
\end{ex}

Substituting (\ref{ss_expr}) into (\ref{pe_hls}), we have
\begin{equation}\label{nume0}
(\sbf{y}-\mbf{S}\mbf{d}-\mbf{R}\mbf{c}-\mbf{Z}\mbf{b})^{T}
(\sbf{y}-\mbf{S}\mbf{d}-\mbf{R}\mbf{c}-\mbf{Z}\mbf{b})
+\mbf{b}^{T}\mbf{\Omega}\mbf{b}+N{\lambda}
\mbf{c}^{T}\mbf{Q}\mbf{c},
\end{equation}
where $\sbf{y}=(\sbf{y}^{T}_1, \cdots, \sbf{y}^{T}_n)^{T}$, $\mbf{d}=(d_1, \cdots, d_m)^{T}$, $\mbf{c}=(c_1, \cdots, c_T)^{T}$,
 $\mbf{b}=(\mbf{b}^{T}_1, \cdots, \mbf{b}^{T}_n)^T$,
$\mbf{S}=(\mbf{S}^{T}_1, \cdots, \mbf{S}^{T}_n)^{T}$ with the
$(k,\nu)$th entry of the $n_i \times{m}$ matrix $\mbf{S}_i$ equal to
$\phi_{\nu}(t_{ik})$, $\mbf{R}=(\mbf{R}^{T}_1, \cdots,
\mbf{R}^{T}_n)^{T}$ with the $(l,j)$th entry of  the $n_i \times{T}$
matrix $\mbf{R}_i$ equal to $R_{M}(t_{il},s_j)$, the design matrix
$\mbf{Z}=\text{diag}(\mbf{Z}_1,\cdots, \mbf{Z}_{n})$,
$\mbf{\Omega}=\sigma^{2}\text{diag}(\mbf{B}^{-1}, \cdots,
\mbf{B}^{-1})$ and $\mbf{Q}$ is $T\times{T}$ matrix with the
$(j,k)$th entry equal to $R_{M}(s_{j},s_{k})$.

Differentiating (\ref{nume0}) with respect to $\mbf{d}$, $\mbf{c}$ and $\mbf{b}$
and setting the derivatives to 0, one has

\begin{equation}\label{norm}
\begin{pmatrix}
\mbf{S}^{T} \mbf{S} & \mbf{S}^{T} \mbf{R} & \mbf{S}^{T} \mbf{Z} \\
\mbf{R}^{T} \mbf{S} & \mbf{R}^{T} \mbf{R} +(N{\lambda} )\mbf{Q} &
\mbf{R}^{T} \mbf{Z}
\\ \mbf{Z}^{T} \mbf{S} & \mbf{Z}^{T} \mbf{R} & \mbf{Z}^{T} \mbf{Z}
+\mbf{\Omega}
\end{pmatrix}
\begin{pmatrix}\hat{\mbf{d}}\\\hat{\mbf{c}}\\\hat{\mbf{b}}\end{pmatrix}
=\begin{pmatrix} \mbf{S}^{T} \sbf{y} \\ \mbf{R}^{T} \sbf{y} \\
\mbf{Z}^{T} \sbf{y}
\end{pmatrix}.
\end{equation}

\noindent The system (\ref{norm}) can be solved through the pivoted
Cholesky decomposition followed by backward and forward
substitutions. See, e.g., \citeN{kg:02} for details.

The fitted values $\hat{\sbf{y}} =\mbf{S} \hat{\mbf{d}} + \mbf{R}
\hat{\mbf{c}}+\mbf{Z}\hat{\mbf{b}}$ of (\ref{pe_hls}) can be written
as $\hat{\sbf{y}} =\mbf{A}(\lambda,\mbf{\Omega})\sbf{y} $, where
$\mbf{A}(\lambda,\mbf{\Omega})$ is the smoothing matrix given below,
\begin{align}\label{smatrix}
\mbf{A} (\lambda,\mbf{\Omega})=(\mbf{S} ,\mbf{R} ,\mbf{Z} )
\begin{pmatrix}
\mbf{S}^{T} \mbf{S} & \mbf{S}^{T} \mbf{R} & \mbf{S}^{T} \mbf{Z} \\
\mbf{R}^{T} \mbf{S} & \mbf{R}^{T} \mbf{R} +(N{\lambda} )\mbf{Q} &
\mbf{R}^{T} \mbf{Z}
\\ \mbf{Z}^{T} \mbf{S} & \mbf{Z}^{T} \mbf{R} & \mbf{Z}^{T}\mbf{Z}
+\mbf{\Omega}
\end{pmatrix}^{+}
\begin{pmatrix}\mbf{S}^{T} \\\mbf{R}^{T} \\\mbf{Z}^{T} \end{pmatrix},
\end{align}

\noindent and $\mbf{C}^{+}$ denotes the Moore-Penrose inverse of
$\mbf{C}$ satisfying $\mbf{C}\mbf{C}^{+}\mbf{C}=\mbf{C}$,
$\mbf{C}^{+}\mbf{C}\mbf{C}^{+}=\mbf{C}^{+}$,
$(\mbf{C}\mbf{C}^{+})^{T}=\mbf{C}\mbf{C}^{+}$ and
$(\mbf{C}^{+}\mbf{C})^{T}=\mbf{C}^{+}\mbf{C}$.

With varying smoothing parameters $\lambda$ (including $\theta$) and
correlation parameters $\mbf{\Omega}$, $(\ref{norm})$ defines an
array of possible estimates, in which we need to choose a specific
one in practice.  A classic data-driven approach for selecting the
smoothing parameter $\lambda$ is generalized cross-validation (GCV),
which was proposed in \citeN{craven:79}. Treating the correlation
parameters $\mbf{\Omega}$ as extra smoothing parameters, we adopt
the approach of \citeN{gm:02} to estimate $\lambda$ and the
correlation parameters $\mbf{\Omega}$ simultaneously through
minimizing the GCV score
\begin{equation}\label{gcv}
V(\lambda,\mbf{\Omega})=\frac{N^{-1}\sbf{y}^{T}(\mbf{I}-\mbf{A}(\lambda,\mbf{\Omega}))^{2}\sbf{y}}
{\{N^{-1}\text{tr}(\mbf{I}-\mbf{A}(\lambda,\mbf{\Omega}))\}^{2}}.
\end{equation}

Since the GCV score $V(\lambda,\mbf{\Omega})$ is non-quadratic in
$\lambda$ and $\mbf{\Omega}$ , one may employ standard nonlinear
optimization algorithms to minimize the GCV as a function of the
tuning parameters. In particular, we used the modified Newton
algorithm developed by \citeN{dennis:96} to find the minimizer. The
distinguishing feature of generalized cross-validation is that its
asymptotic optimality can be justified in a decision-theoretic
framework. One may define a quadratic loss function as,
$$L(\lambda,\mbf{\Omega})=\frac{1}{N}\sum_{i=1}^{n}(\hat{\sbf{y}}_{i}-\mu(\mbf{x}_i)-\mbf{Z}_i\mbf{b}_i)^{T}(\hat{\sbf{y}}_{i}-\mu(\mbf{x}_i)-\mbf{Z}_i\mbf{b}_i).$$

Under general conditions, \citeN{gm:02} showed that the GCV tracks
the loss function asymptotically,
\[
V(\lambda,\mbf{\Omega})-L(\lambda,\mbf{\Omega})-\frac{1}{N}\sum_{i=1}^{n}
\sbf{\epsilon}_{i}^{T}\sbf{\epsilon}_{i}
=o_{p}(L(\lambda,\mbf{\Omega})).
\]

Note that $\sbf{\epsilon}_{i}$ does not depend on $\lambda$ and
$\mbf{\Omega}$. It then follows that the minimizer of the GCV score
$V(\lambda,\mbf{\Omega})$ approximately minimizes the loss function
$L(\lambda,\mbf{\Omega})$.

\subsection{Bayesian Confidence Intervals}

Unlike confidence estimates in parametric models, a rigorously
justified interval estimate is a rarity for nonparametric functional
estimation.  An exception is the Bayesian confidence interval
developed by \citeN{wahba:83} from a Bayes model. A nice feature of
Bayesian confidence intervals is that they have a certain
across-the-function coverage property. See \citeN{nychka:88}. In
this section, we derive the posterior mean and variance for
constructing Bayesian confidence intervals in our setting.

The regularization is equivalent to imposing a prior on the
functional form of $\mu(x)$. To see this,we decompose $\mu=f_0+f_1$,
where $f_0$ has a diffuse prior in the space $\mathcal{N}_{M}$ and
 $f_1$ has an independent Gaussian process prior with mean zero and covariance,
\begin{equation}\label{ppp}
E[f_1(s_{k})f_1(s_{l})] =\frac{\sigma^2}{N{\lambda}
}R_{M}(s_{k},\mbf{s}^{T})\mbf{Q}^{+}R_{M}(\mbf{s},s_{l}).
\end{equation}
The minimizer of (\ref{pe_hls}) can be shown to be the posterior
mean under the above prior by the following theorem.

\begin{thm}\label{thm1}
With the prior for $\mu$ specified above and a generic $np \times 1$
vector $\mbf{z}$, the posterior mean of $\mu(x)+\mbf{z}^T\mbf{b}$
has the following expression:
\begin{equation}\label{post_m}
E[\mu(x)+\mbf{z}^{T}\mbf{b}|\sbf{y}]=\sbf{\phi}^T\hat{\mbf{d}} +
\sbf{\xi}^T\hat{\mbf{c}} +\mbf{z}^{T}\hat{\mbf{b}},
\end{equation}
where $\sbf{\phi}$ is $m\times 1$ with the $\nu$th entry
$\phi_{\nu}$(x), $\sbf{\xi}$ is $T \times 1$ with the $i$th entry
$R(s_i,x)$, $\hat{\mbf{d}}$, $\hat{\mbf{c}}$, and $\hat{\mbf{b}}$
are the solutions of (\ref{norm}).
\end{thm}

The posterior variance is given in the following theorem.
\begin{thm}\label{thm2}
Under the model specified in Theorem \ref{thm1}, the posterior
variance has the following expression:
\begin{multline}
\frac{N{\lambda}
}{\sigma^2}\text{Var}[\mu(x)+\mbf{z}^{T}\mbf{b}|\sbf{y}]
=\sbf{\xi}^{T}\mbf{Q}^{+}\sbf{\xi}+N{\lambda}  \mbf{z}^{T}\mbf{\Omega}^{+}\mbf{z}+\sbf{\phi}^{T}(\mbf{S}^{T}\mbf{W}^{-1}\mbf{S})^{-1}\sbf{\phi}\notag\\
\quad-2\sbf{\phi}^{T}(\mbf{S}^{T}\mbf{W}^{-1}\mbf{S})^{-1}\mbf{S}^{T}\mbf{W}^{-1}\mbf{R}\mbf{Q}^{+}\sbf{\xi}-2N{\lambda} \sbf{\phi}^{T}(\mbf{S}^{T}\mbf{W}^{-1}\mbf{S})^{-1}\mbf{S}^{T}\mbf{W}^{-1}\mbf{Z}\mbf{\Omega}^{+}\mbf{z}\notag\\
\quad-(\sbf{\xi}^{T}\mbf{Q}^{+}\mbf{R}^{T}+N{\lambda}
\mbf{z}^{T}\mbf{\Omega}^{+}\mbf{Z})(\mbf{W}^{-1}-\mbf{W}^{-1}\mbf{S}(\mbf{S}^{T}\mbf{W}^{-1}\mbf{S})^{-1}\mbf{S}^{T}\mbf{W}^{-1})(\mbf{R}\mbf{Q}^{+}\sbf{\xi}+N{\lambda}
\mbf{Z}\mbf{\Omega}^{+}\mbf{z}),\label{post_v}
\end{multline}

where $\mbf{W}=\mbf{R}\mbf{Q}^{+}\mbf{R}^{T}+N{\lambda}
\mbf{Z}\mbf{\Omega}^{+}\mbf{Z}^{T}+N{\lambda} \mbf{I}$.
\end{thm}

The proofs of the above two theorems are given in Appendix. Using
Theorem \ref{thm1} and Theorem \ref{thm2}, we construct the
$100(1-\alpha)\%$ Bayesian confidence intervals as,
$E[\mu(x)+\mbf{z}^{T}\mbf{b}|\sbf{y}] \pm
\Phi(1-\alpha/2)^{-1}\sqrt{\text{Var}[\mu(x)+\mbf{z}^{T}\mbf{b}|\sbf{y}]}$,
 where $\Phi(1-\alpha/2)^{-1}$ is the $100(1-\alpha/2)$ percentile of the standard Gaussian
 distribution. Letting $\mbf{z}=0$, we get Bayesian confidence intervals
 for $\mu(x)$.  Note that the construction
of Bayesian confidence intervals is pointwise. It is unclear whether
the across-the-function coverage property of \citeN{nychka:88} holds
in our case.

\section{The Mixture Model}

Based on the mixed-effect representation of homogeneous functional
data, we shall now present a mixture model for characterizing the
heterogeneity.

\subsection{The Model Specification}

When the population is heterogeneous, we assume that the $i$th
functional observation can be modeled as
\begin{equation}\label{mmixedn}
\sbf{y}_i= \mu_k(\mbf{x}_i)+ \mbf{Z}_i\mbf{b}_i+ \sbf{\epsilon}_i
\quad\quad \text{with probability $p_k$}
\end{equation}
where $k=1, \cdots, K$, the $k$th cluster's mean $\mu_k$ is a smooth
function defined on a generic domain $\Gamma$,
$\mbf{b}_i\sim{N}(0,\mbf{B}_{k})$ is a $p\times{1}$ random effect
vector associated with a $n_i\times p$ design matrix $\mbf{Z}_i$,
$\sbf{\epsilon}_i\sim{N}(0,\sigma^{2}\mbf{I})$ are random errors
independent of the $\mbf{b}_i$'s and of each other, cluster
probabilities $p_k$ satisfy $\sum_{k=1}^{K}p_k=1$, and $K$ is the
number of clusters in the population.

To ease the computation, we introduce a ``latent'' membership
labeling variable $J_{ik}$ such that $J_{ik}=1$ indicates individual
$i$ belongs to the $k$th cluster and $J_{ik}=0$ otherwise. Thus we
have the probability that $J_{ik}=1$ is $p_k$. The mixture
Henderson's likelihood is seen to be
$$\sum_{i=1}^{n}\log{\sum_{k=1}^{K}[p_{k} f_y(\sbf{y}_i; \mbf{b}_i, J_{ik}=1)f_b(\mbf{b}_i ; J_{ik}=1)]}$$
where $f_y$ and $f_b$ are probability density functions for
$\sbf{y}_i$ and $\mbf{b}_i$ respectively.

\subsection{Estimation}

The negative penalized Henderson's likelihood of complete data
$(\sbf{y}_i, J_{ik})$ where $i=1, \cdots, n$, is seen to be
\begin{multline}\label{cl}
L_c=\text{Constant} -\sum_{i=1}^{n}\sum_{k=1}^{K}J_{ik}\log{p}_k\\
+\frac{1}{2\sigma^2}\sum_{i=1}^{n}\sum_{k=1}^{K}J_{ik}[(\sbf{y}_i-\mu_k(\mbf{x}_i)-\mbf{Z}_i\mbf{b}_i)^{T}(\sbf{y}_i-\mu_k(\mbf{x}_i)-\mbf{Z}_i\mbf{b}_i)
+\sigma^2\mbf{b}^{T}_i \mbf{B}_k^{-1} \mbf{b}_i] +
\sum_{k=1}^{K}{N}\lambda_k M(\mu_k)
\end{multline}
where $\lambda_k$ is the smoothing parameter for  $\mu_k$.

Once the penalized Henderson's likelihood (\ref{cl}) is obtained,
the EM algorithm (\shortciteNP{em:77}, \citeNP{Green:1990}) can be
derived as follows.

The E-step simply requires the calculation of
\begin{align}\label{tk}
w_{ik}=\frac{p_{k}\varphi(\sbf{y}_i;\mu_k(\mbf{x}_i),
\mbf{\Sigma}_k)}{\sum_{l=1}^{K}p_{l}\varphi(\sbf{y}_i;\mu_l(\mbf{x}_i),
\mbf{\Sigma}_l)}
\end{align}
where
$\mbf{\Sigma}_k=\mbf{Z}_i\mbf{B}_k\mbf{Z}_i^{T}+\sigma^2\mbf{I}$,
and $\varphi$ is the Gaussian density function.

The M-step requires the conditional minimization of the following
equation
\begin{align}
&-\sum_{k=1}^{K}\sum_{i=1}^{n}w_{ik}\log{p}_k\label{E1}\\
&+\frac{1}{2\sigma^2}\sum_{i=1}^{n}\sum_{k=1}^{K}w_{ik}[(\sbf{y}_i-\mu_k(\mbf{x}_i)-\mbf{Z}_i\mbf{b}_{ik})^{T}(\sbf{y}_i-\mu_k(\mbf{x}_i)-\mbf{Z}_i\mbf{b}_{ik})
+\sigma^2\mbf{b}_{ik}^{T} \mbf{B}_{k}^{-1} \mbf{b}_{ik}]+
\sum_{k=1}^{K}N \lambda_k M(\mu_k),\label{E2}
\end{align}
where $\mbf{b}_{ik}$ is $\mbf{b}_{i}$ given the membership $J_{ik}$.
Thus the M-step is equivalent to minimizing (\ref{E1}) and
(\ref{E2}) separately.

By minimizing (\ref{E1}), we have
\begin{equation}\label{p-est}
p_k=\frac{1}{n}\sum_{i=1}^n w_{ik} \quad \text{for}\quad k= 1, \cdots,  K.
\end{equation}

By minimizing  (\ref{E2}), we can minimize the following $K$
equations simultaneously
\begin{equation}\label{multE2}
\sum_{i=1}^{n}
w_{ik}[(\sbf{y}_i-\mu_k(\mbf{x}_i)-\mbf{Z}_i\mbf{b}_{ik})^{T}(\sbf{y}_i-\mu_k(\mbf{x}_i)-\mbf{Z}_i\mbf{b}_{ik})
+\sigma^2\mbf{b}^{T}_{ik} \mbf{B}_k^{-1} \mbf{b}_{ik}]+ N\lambda_k
M(\mu_k)\quad k= 1, \cdots,  K.
\end{equation}
where $1/{2\sigma^2}$ is absorbed into $\lambda_k$. The minimization
of (\ref{multE2}) is performed in the reproducing kernel Hilbert
space $\mathcal{H}\subseteq\{\eta:M(\mu)<\infty\}$. Substituting
solution (\ref{ss_expr}) into (\ref{multE2}), we have
\begin{equation}\label{nume0w}
(\sbf{y}-\mbf{S}\mbf{d}_k-\mbf{R}\mbf{c}_k-\mbf{Z}\mbf{b}_k)^{T}\mbf{W}_k
(\sbf{y}-\mbf{S}\mbf{d}_k-\mbf{R}\mbf{c}_k-\mbf{Z}\mbf{b}_k)
+\mbf{b}_k^{T}\tilde{\mbf{W}}_k^{1/2}\mbf{\Omega}_k\tilde{\mbf{W}}_k^{1/2}\mbf{b}_k+N{\lambda_k}
\mbf{c}_k^{T}\mbf{Q}\mbf{c}_k,
\end{equation}
where $\sbf{y}=(\sbf{y}^{T}_1, \cdots, \sbf{y}^{T}_n)^{T}$, $\mbf{d}_k=(d_{1k}, \cdots, d_{mk})^{T}$, $\mbf{c}_k=(c_{1k}, \cdots, c_{Tk})^{T}$,
 $\mbf{b}_k=(\mbf{b}^{T}_{1k}, \cdots, \mbf{b}^{T}_{nk})^T$,
$\mbf{S}=(\mbf{S}^{T}_1, \cdots, \mbf{S}^{T}_n)^{T}$ with the
$(k,\nu)$th entry of the $n_i \times{m}$ matrix $\mbf{S}_i$ equal to
$\phi_{\nu}(t_{ik})$, $\mbf{R}=(\mbf{R}^{T}_1, \cdots,
\mbf{R}^{T}_n)^{T}$ with the $(l,j)$th entry of  the $n_i \times{T}$
matrix $\mbf{R}_i$ equal to $R_{M}(t_{il},s_j)$, the design matrix
$\mbf{Z}=\text{diag}(\mbf{Z}_1,\cdots, \mbf{Z}_{n})$,
$\mbf{W}_k=\text{diag}(w_{1k}\mbf{I}_{n_1}, \cdots,
w_{nk}\mbf{I}_{n_n})$, $\tilde{\mbf{W}}_k=\text{diag}(w_{1k}
\mbf{I}_{p}, \cdots, w_{nk} \mbf{I}_{p})$,
$\mbf{\Omega}_k=\sigma^{2}\text{diag}(\mbf{B}_k^{-1}, \cdots,
\mbf{B}_k^{-1})$ and $\mbf{Q}$ is the $T\times{T}$ matrix with the
$(j,k)$th entry equal to $R_{M}(s_{j},s_{k})$.

Writing (\ref{nume0w}) in a more compact form, we have
\begin{equation}\label{numew}
(\sbf{y}_{wk}-\mbf{S}_{wk}\mbf{d}_k-\mbf{R}_{wk}\mbf{c}_k-\mbf{Z}\mbf{b}_{wk})^{T}
(\sbf{y}_{wk}-\mbf{S}_{wk}\mbf{d}_k-\mbf{R}_{wk}\mbf{c}_k-\mbf{Z}\mbf{b}_{wk})
+\mbf{b}^{T}_{wk}\mbf{\Omega}_k\mbf{b}_{wk}+N{\lambda}_k
\mbf{c}_k^{T}\mbf{Q}\mbf{c}_k,
\end{equation}
where $\sbf{y}_{wk}=\mbf{W}_k^{1/2}\sbf{y}$,
$\mbf{S}_{wk}=\mbf{W}_k^{1/2}\mbf{S}$,
$\mbf{R}_{wk}=\mbf{W}_k^{1/2}\mbf{R}$,
$\mbf{Z}_{wk}=\mbf{W}_k^{1/2}\mbf{Z}\tilde{\mbf{W}}_k^{-1/2}$, and
$\mbf{b}_{wk}=\tilde{\mbf{W}}_k^{1/2}\mbf{b}_k$. Then (\ref{numew})
can be minimized using the techniques developed in Section 2.

The variance of measurement error is estimated as
\begin{equation}\label{sigma2-mix}
\hat{\sigma}^2=\frac{1}{N}\sum_{i=1}^{n}\sum_{k=1}^{K}w_{ik}(\sbf{y}_i-\mu_k(\mbf{x}_i)-\mbf{Z}_i\mbf{b}_{ik})^{T}(\sbf{y}_i-\mu_k(\mbf{x}_i)-\mbf{Z}_i\mbf{b}_{ik}).
\end{equation}

The algorithm iterates through (\ref{tk}), (\ref{p-est}),
(\ref{numew}) and (\ref{sigma2-mix}) until all the parameters
converge.

The selection of the smoothing parameters $\mbf{\Omega}_k$ and
$\lambda_k$ plays an important role in the proposed algorithm. When
first run our algorithm, in each iteration, the optimal smoothing
parameters are selected for each cluster using GCV in (\ref{numew}).
Once all parameters converge, we fixed the selected smoothing
parameters and run our algorithm for fixed smoothing parameters.

After we fit the mixture model to the data, we can give a
probabilistic (soft) clustering of each observation $\sbf{y}_i$.
That is, for each $\sbf{y}_i$, $w_{i1},\cdots, w_{iK}$ give the
estimated probabilities that this observation belongs to the first,
second,..., and $K$th components, respectively, of the mixture.
However, in many practical settings, it is highly desirable to give
hard clustering of these observations by assigning each observation
to one component of the mixture. In the rest of the paper, we adopt
the hard clustering of \citeN{McLa:Peel:fini:2001} by estimating the
membership label,
\begin{displaymath}
\hat{J}_{ik}=\left\{\begin{array}{l}1 \quad\text{if}\quad k=\text{argmax}_{h}w_{ih}\\
               0 \quad \text{otherwise}
\end{array}\right.
\end{displaymath}
where $k=1, \cdots, K$ and $i=1, \cdots, n$.

\subsection{Efficient Computation with Rejection Control}

 With thousands of observations under consideration, the E-step (\ref{tk}) results
 in a huge number of $w_{ik}$'s, many of which are extremely small.
 With the presence of
 these small $w_{ik}$'s, the calculation of matrices involved in the M-step
 (\ref{numew}) is expensive, unstable and sometimes even infeasible. To alleviate
the computation and stabilize the algorithm, we propose to add a
rejection control step (\shortciteNP{Liu:Chen:Wong:1998}) in the EM
algorithm and refer to the modified algorithm as rejection
controlled EM algorithm.

Firstly, we set up a threshold value $c$ (e.g., $c=0.05$). Given
this threshold value, we introduce the following rejection
controlled step:
\begin{displaymath}
w_{ik}^{*}=\left\{\begin{array}{ll}\label{rejection}
w_{ik}  \quad \text{if}\quad  w_{ik}>c\\
c \quad  \text{with probability} \quad w_{ik}/c \quad \text{if} \quad  w_{ik} \le c\\
0 \quad \text{with probability}\quad  1-w_{ik}/c \quad \text{if}
\quad w_{ik} \le c.
\end{array}\right.
\end{displaymath}
The resulting $w_{ik}^{*}$ needs to be normalized:
$w_{ik}^{**}=w_{ik}^{*}/\sum_{k}w_{ik}^{*}$. Then we replace
$w_{ik}$ by $w_{ik}^{**}$ right after the E-step (\ref{tk}). Note
that when $c=0$, the proposed algorithm is exactly the original EM
algorithm, whereas the proposed algorithm reduces to a variant of
Monte Carlo EM algorithm (\citeNP{Wei:Tann:1990}) when $c=1$. In
this way, it is possible to make accurate approximations during the
E-step while greatly reducing the computation of the M-step.

Finally, in order to avoid local optima, the rejection controlled EM
is run with multiple chains. In practice, we first set the threshold
$c$ close to 1 at an early stage of the iterations to expedite the
calculation, then we gradually lower $c$ so that the algorithm can
achieve a better approximation of the original EM.

A critical issue arising from the new algorithm is how to choose an
appropriate stopping rule. For the original EM algorithm, the
likelihood function increases after each iteration, so we can stop
the iteration when the likelihood does not change. However, for the
rejection controlled EM algorithm, the likelihood functions
fluctuates because of the sampling scheme. So a stopping rule like
those used in the Gibbs sampler is employed. When the likelihood
function is no longer increasing for several consecutive iterations,
we stop and choose the estimates with the highest likelihood.

\subsection{The Selection of the Number of Clusters}

The success of our proposed methods heavily depends on the selection
of the number of clusters $K$. A natural choice in model-based
clustering is to use the Bayesian Information Criterion (BIC). The
BIC imposes a penalty on the total number of parameters, scaled by
the logarithm of sample size, so as to strike a balance between the
goodness-of-fit and the model complexity.  A critical issue in using
BIC in nonparametric settings is to determine the effective number
of parameters. Here we use the trace of the smoothing matrix to
approximate the number of parameters in each cluster
(\citeNP{ht:90}, \citeNP{gu:02}). Thus BIC under our model is
\begin{equation}\label{bic}
BIC=-2\sum_{i=1}^{n}\log\sum_{k=1}^K
p_{k}\varphi(\sbf{y}_i;\mu_k(\mbf{x}_i),
\mbf{\Sigma}_k)+(\sum_{k=1}^{K}\text{tr}\mbf{A}_{k}(\lambda_k,\mbf{\Omega}_k)
+ P)\log{N},
\end{equation}
where $\mbf{A}_k$ is the smoothing matrix for the $k$th cluster as
defined in (\ref{smatrix}), $P$ is the number of free parameters in
$p_k$, $\lambda_k$, and $\mbf{\Omega}_k$ where $k=1, \cdots, K$.

\section{Simulation}
To assess the performance of the proposed method, we carried out
extensive analysis on simulated datasets.

This simulation is designed to demonstrate the performance of the
proposed method when the underlying clusters' mean functions are
different for different clusters.
 First, one hundred replicates
of samples were generated according to
\begin{align*}
y_{1ij\tau}&=3 \sin(6 \pi t_j)(1-t_j)+ 2I_{\{1\}}(\tau)-1 + \epsilon_{1ij\tau}, \quad i=1,\cdots,30; \\
y_{2ij\tau}&=3 \sin(6 \pi t_j) (1-t_j)+\epsilon_{2ij\tau}, \quad i=1,\cdots,40;\\
y_{3ij\tau}&=1980 t_j^7 (1-t_j)^3+ 858 t_j^2 (1-t_j)^{10} - 2 + \epsilon_{3ij\tau}, \quad i=1,\cdots,50;\\
y_{4ij\tau}&=3 \sin(2 \pi t_j)+ 2 I_{\{1\}}(\tau)-1 +
\epsilon_{4ij\tau}, \quad i=1,\cdots,30;
\end{align*}
where $t_j=1/15, 2/15, \cdots, 1$, $\tau=0, 1$, indicator function
$I_{\{1\}}(\tau)=1$ if $\tau=1$ and 0 otherwise,
 random errors $\epsilon$ were
generated from a Gaussian distribution with mean zero and covariance
matrix as follows:
\begin{align*}
\text{Var}[\epsilon_{lij\tau}]&=1, \quad\quad \text{Cov}(\epsilon_{lij\tau_1}, \epsilon_{lik\tau_2})=0.2, \quad\quad \text{for}\quad  l=1,3; \\
\text{Var}[\epsilon_{lij\tau}]&=1.2, \quad\quad
\text{Cov}(\epsilon_{lij\tau_1}, \epsilon_{lik\tau_2})=0.4,
\quad\quad \text{for}\quad l=2,4;
\end{align*}

We analyzed the simulated data using the proposed method with the
following mixture model
$$\sbf{y}_i= \mu_k(\mbf{t}, \tau)+ b_i \mbf{1}+ \sbf{\epsilon}_i
\quad\quad \text{with probability $p_k$},$$ where $k=1, \cdots, K$,
$\tau=1, 2$ for two groups, $b_i \sim N(0, \sigma^2_b)$ is the
individual specific random effect. The important feature of the
simulated data is that the true mean curves in two groups, indexed
by $\tau$, are either identical or parallel. This information was
built into our method through enforcing the additive model
(\ref{addspline}). The penalized Henderson's likelihood was employed
for estimation with roughness penalty
$M(\mu)=\int_{0}^{1}(d^{2}\mu_{1}/dt^{2})^{2}dt$.
\begin{figure}
\centerline{\psfig{file=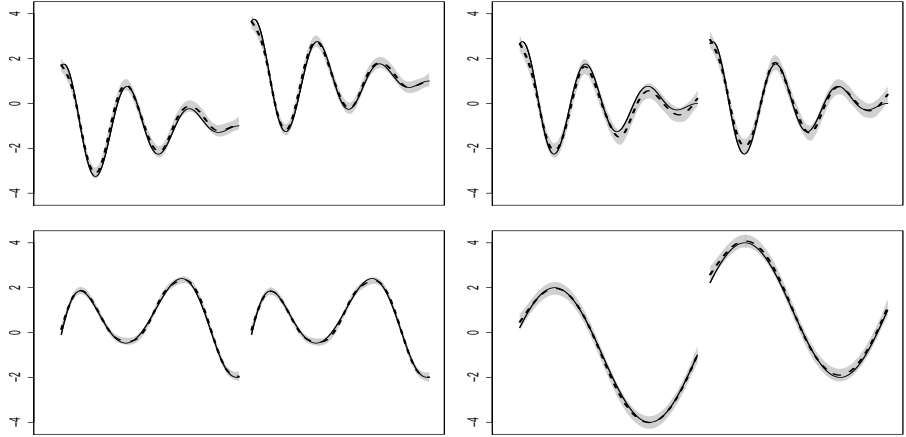,height=.4\linewidth,width=0.8\linewidth,clip=,angle=270}}
\caption{The estimated mean curves (dash lines) and 95\% Bayesian
confidence intervals for one simulated dataset. The true functions
are superimposed as solid lines.} \label{simu:fig}
\end{figure}

We compared  our method with MCLUST (\citeNP{fraley:2002}),  FCM
classification likelihood (FCMc), and FCM mixture likelihood (FCMm)
(\citeNP{Jame:Suga:2003}). Since the number of clusters must be
specified {\it a priori} in the partially implemented FCM software,
we gave a significant starting advantage to the FCM algorithm by
letting the number of clusters be the true number of clusters
(four). For MCLUST, the clustering result with optimal BIC was
reported, which was estimated from eight models with different
covariance structures. The estimated mean curves using the proposed
method for each cluster and the true curves of one sample are
plotted in Figure~\ref{simu:fig}.

For comparison, we need a measure of the agreement of the clustering
results with the true cluster membership. A popular one is the Rand
index, which is the percentage of concordance pairs over all
possible data pairs. \citeN{huber:1985} proposed an adjusted Rand
index, which takes one as the maximum value when two clustering
results are the same and the expected value is equal to zero when
two clustering results are independent.  We found that across 100
samples the average of the adjusted Rand indices for the proposed
method is 0.9676 (median is 0.9838), whereas those of MCLUST, FCMm,
FCMc are 0.7553, 0.8936, and 0.8896, respectively. Moreover, the
inter-quartile range of the adjusted Rand indices of the proposed
method is 0.0565, (0.0972, 0.2189 and 0.2262 for MCLUST, FCMm, and
FCMc respectively). These results suggest that the proposed method
outperforms FCMc and FCMm (even under the ideal scenario where the
true number of clusters is provided to FCMc and FCMm a priori) as
well as MCLUST.

\section{Real Data Examples}

\subsection{Comparative Genomic Study of Fruitfly and Worm Gene Expressions}

Development is an important biological process that shares many
common features among different organisms. It is well-known that
{\it D. melanogaster} (fruitfly) and {\it C. elegance} (worm) are
two highly diverged species, the last common ancestor of which
existed about one billion years ago. Their development is an active
research area: In \shortciteN{arbeitman:02}, the mRNA levels of 4028
genes in {\it D. melanogaster} were measured using cDNA microarrays
during 62 time points starting at fertilization and spanning
embryonic, larval, pupal (metamophosis) stages and the first 30 days
of adulthood. mRNA was extracted from mixed male and female
populations until adulthood when males and females were sampled
separately. \shortciteN{jiang:01} reported a cDNA microarray
experiment for 17871 genes over the life-cycle of {\it C. elegans}
at 6 time points, including eggs, larval stages: L1, L2, L3 and L4,
and young adults.

\begin{figure}
\centerline{\psfig{file=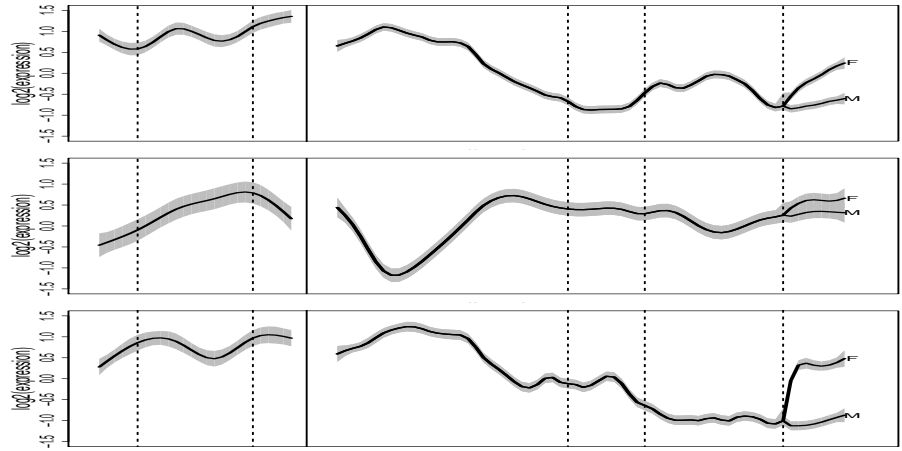,height=0.4\linewidth,width=0.8\linewidth,clip=,angle=270}}
\caption{Estimated mean expression curves and $95\%$ Bayesian
confidence intervals (grey bands) for cluster A, B and C (from top
to bottom) discovered in the worm-fly temporal expression data.
Vertical solid lines separate worm (eggs, larva and young adult are
separately by dash lines in the left frame), fruit-fly
(embryogenesis, larva, pupa, and adult stages are separated by dash
lines in the right frame). Adult fruit-fly male and female mean
expression curves are labeled as M and F, respectively.}
\label{fly:fig}
\end{figure}

To study the genomic connections in expression patterns across the
two species, we combined the gene expression datasets of
\shortciteN{arbeitman:02} and  \shortciteN{jiang:01} using the
orthologous genes provided by \shortciteN{haoli:04}, which resulted
in a merged expression dataset containing 808 orthologous genes. We
analyzed the data using the proposed method with the mixture model,
$$\sbf{y}_i= \mu_k(\mbf{t}, \tau)+ b_i \mbf{1}+ \sbf{\epsilon}_i$$
with probability $p_k$ where $k=1, \cdots, K$, $\tau=1$ for fruitfly
and $\tau=2$ for worm, $b_i \sim N(0, \sigma^2_b)$ is the gene
specific random effect. The penalized Henderson's likelihood was
employed with roughness penalty $M$ of the form (\ref{mixpenal}).
 Sex differentiation of the fruitfly was modeled by a branching spline
(\citeNP{silv:wood:1987}), the general analytic form of which with
two branches on the right is
\begin{equation}\notag
\mu(t)=\left\{\begin{array}{l}
\sum_{\nu=1}^{m}d_{\nu}\phi_{\nu}(t)+\sum_{i=1}^{k}{c}_{i}R_{M}(s_{i},t)\quad \text{if}\quad t \leq s_k \\
 \sum_{\nu=1}^{m}d_{\nu}\phi_{\nu}(t)+\sum_{i=1}^{k}{c}_{i}R_{M}(s_{i},t)  +\sum_{i=k+1}^{T}{c}_{1i}R_{M}(s_{i}-s_k,t-s_k) \quad \text{if} \quad t > s_k \\
 \sum_{\nu=1}^{m}d_{\nu}\phi_{\nu}(t)+\sum_{i=1}^{k}{c}_{i}R_{M}(s_{i},t)
 +\sum_{i=k+1}^{T}{c}_{2i}R_{M}(s_{i}-s_k,t-s_k) \quad \text{if}\quad t >
 s_k
\end{array}\right.
\end{equation}
where $s_k$ is the branching point, and the second and third rows
are expressions of the two branches. A cubic smoothing spline was
used. The 808 genes were clustered by our method into 34 clusters.
Biological functions of genes in each cluster were annotated using
Gene Ontology, and Bonferroni corrected P-values of biological
function enrichment were calculated based on the hypergeometric
distribution (\citeNP{cristian:03}). Of the 34 clusters discovered,
21 clusters exhibit significant biological functions
over-representation (P-value $<$ 0.05). The estimated mean gene
expression curves of three clusters and their 95\% Bayesian
confidence intervals are given in Figure~\ref{fly:fig}.

In cluster A, which consists of 31 genes, gene expressions of worms
have peaks at eggs, larva and young adult. In the same cluster, we
observed that fruit-fly gene expressions that are up-regulated
during embryogenesis are also up-regulated during metamorphosis,
suggesting that many genes used for pattern formation during
embryogenesis (the transition from egg to larva) are re-deployed
during metamorphosis (the transition from larva to adult).
Consistently, this cluster is enriched for genes involved in
embryonic development (P-value =0.0003),  post-embryonic body
morphogenesis (P-value =0.007), and mRNA processing (P-value =
0.002), among others.

In cluster B, consisting of 24 genes, gene expressions of worms
increase starting at eggs until they reach a peak at late larval
stage. Then expressions go down during adulthood. However, we
observed that fruit-fly gene expressions that are down-regulated
during embryogenesis are up-regulated during metamorphosis and
adult, suggesting that many genes are involved in development. The
enriched gene functions are embryonic (P-value =0.02), larval
development (P-value =0.008), and growth regulation (P-value $<
10^{-5}$ ).

Cluster C contains 25 genes. For worms, gene expressions have peaks
at larva and adult stages. An over-representation of gene functions
such as reproduction (P-value $< 10^{-6}$), larval development
(P-value $< 10^{-7}$). Fruit-flies show peaks in gene expression in
the early embryo, and older females (but not males). An
over-representation of gene functions such as reproduction (P-value
$< 10^{-6}$) and embryonic development(P-value $< 10^{-5}$) are
present in this cluster. Among related functions, this cluster also
contains functions of female gamete generation, growth, and positive
regulation of growth rate. Genes of this cluster are thus inferred
to participate in sex determination, female production of eggs, and
growth regulation.

\subsection{Budding Yeast Gene Expression under Aerobic and Anaerobic Conditions}

To study the oxygen-responsive gene networks, \shortciteN{lai:2006}
used cDNA microarray to monitor the gene expression changes of
wild-type budding yeast ({\it Saccharomyces cerevisiae}) under
aerobic and anaerobic conditions in a galactose medium. Under the
aerobic conditions, the oxygen concentration was lowered gradually
until oxygen was exhausted during a period of ten minutes. After 24
hours of anaerobiosis, the oxygen concentration was progressively
increased back to normal level during another period of ten minutes,
which was referred to as the anaerobic conditions. Microarray
experiments were conducted at 14 time points under aerobic
conditions and 10 time points under anaerobic conditions. A
reference sample pooled from all time points was used for
hybridization.
\begin{figure}
\centerline{\psfig{file=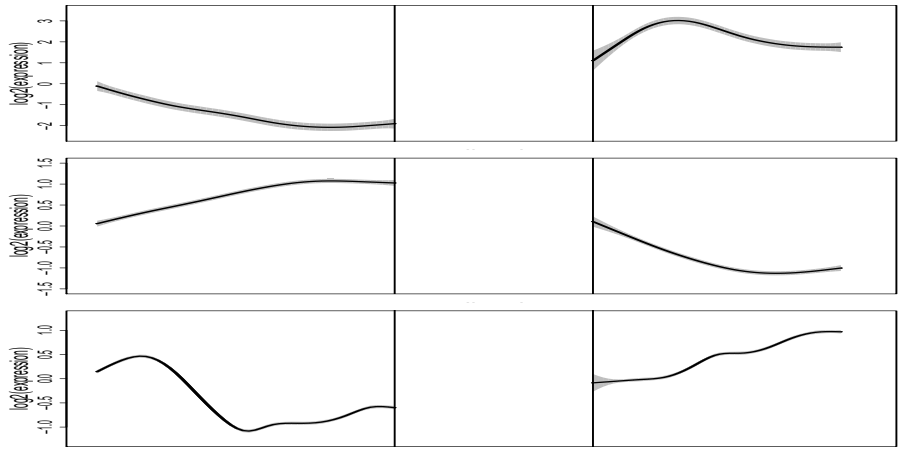,height=0.4\linewidth,width=0.8\linewidth,clip=,angle=270}}
\caption{Estimated mean expression curves and $95\%$ Bayesian
confidence intervals (grey bands) for cluster A, B and C (from top
to bottom) discovered in the yeast aerobic and anaerobic expression
data. The aerobic (left) and anaerobic (right) conditions were
separated by two vertical lines.} \label{yeast:fig}
\end{figure}

For their analysis, \shortciteN{lai:2006} normalized gene
expressions to gene expressions of time 0 and filtered out
differentially expressed genes. Thus the normalized expressions at
23 time points of 2388 differentially expressed genes are used for
our clustering analysis. We modeled normalized gene expression
$\sbf{y}_i$ of the $i$th gene using the mixture model, $$\sbf{y}_i=
\mu_k(\mbf{t}, \tau)+ b_i \mbf{1}+ \sbf{\epsilon}_i \quad \text{with
probability $p_k$}$$ where $k=1, \cdots, K$, $\tau=1$ for aerobic
and $\tau=2$ for anaerobic condition, $b_i \sim N(0, \sigma^2_b)$ is
the gene specific random effect. We fit the model using the penalty
(\ref{mixpenal}) with $a=2$. In total, 2388 genes were clustered
into 28 clusters using our method.  FunSpec
(\shortciteNP{funspec:2002}) was used for gene annotation and
biological function enrichment analysis. We found 26 clusters out of
28 clusters discovered have over-represented biological functions.
The estimated mean gene expression profiles and associated Bayesian
confidence intervals of three clusters are given in
Figure~\ref{yeast:fig}.

In cluster A, which consists of 57 genes, the estimated mean
expression goes down progressively as oxygen level goes down, which
suggests that the genes in this cluster are transiently
down-regulated in response to anaerobisis. Furthermore, the
estimated mean expression increases as oxygen concentration shifts
back to normal level. Accordingly, genes involved in respiration,
lipid fatty-acid and isoprenoid biosynthesis, and cell defense are
over-represented in this cluster (P-value $\le 10^{-5}$).

In contrast to cluster A, cluster B (85 genes) consists of genes
involved in various biosynthesis, metabolism and catabolism such as
glucose metabolism (P-value$\le 10^{-6}$).   These biological
processes are necessary to maintain the basic living needs of yeast
cells. Interestingly, the alcohol biosythesis and metabolism are
also enriched in this cluster. Consistent with biological function
over-representation, the estimated mean expression is up-regulated
in aerobic conditions and down-regulated in anaerobic conditions.

We have 70 genes in cluster C, where the estimated mean gene
expression goes up at the beginning and then drops down rapidly
under aerobic conditions. Under anaerobic conditions, the estimated
mean gene expression is up-regulated. In this cluster, respiratory
deficiency and carbon utilization are also over-represented
(P-value$\le 10^{-8}$). The initial up-regulation of gene expression
under aerobic conditions can be partly explained by the fact that
the cell increases energy up-taking through other biological
processes, such as carbon utilization, when oxygen goes down. But as
the oxygen level continues to drop, these processes are replaced by
more energy efficient processes, such as glucose metabolism. Under
the anaerobic conditions, these processes are revitalized again as
oxygen level increases.

\section{Discussion}

In this article, we propose a clustering method for large scale
functional data with multiple covariates. Nonparametric mixed-effect
models were built, which were nested under a mixture model. The
penalized Henderson's likelihood was employed for estimation.
Data-driven smoothing parameters, selected through generalized
cross-validation, were used to automatically capture the functional
features. The rejection-controlled EM algorithm was designed to
reduce the expensive computational cost for large scale data. The
simulation analyses suggest that the proposed method outperforms the
existing clustering methods. Moreover, the Bayesian interpretation
of the proposed method allows the development of an equivalent fully
Bayesian functional data clustering method, which can accommodate
additional genomic and proteomic information for gene expression
study. Although it  was motivated for clustering temporal expression
data, our proposed method has a wide spectrum of applications,
including those involving seismic wave data arising from geophysical
research (\shortciteNP{Wang:2006} and \shortciteNP{Ma:2007}). The
calculations reported in this article were performed in R.
Open-source code is available in the R package MFDA.

As a sequel to this work, a clustering method for discrete data,
especially those arising from temporal text mining, is under active
development.

\section*{Appendix}

{\it Proof of Theorem \ref{thm1} :}

Note the fact that if we specify the prior for $f_0$ as a Gaussian
process with mean zero and covariances
$E[f_0(s_{k})f_0(s_{l})]=\tau^2 \sum_{\nu = 1}^{m}
\phi_{\nu}(s_{k})\phi_{\nu}(s_{l})$, then when
$\tau^{2}\rightarrow\infty$, the prior for $f_0$ becomes a diffuse
prior; see \citeN{wahba:83} and \citeN{gu:02}.

Assuming $f_0(t)$ has a Gaussian process prior specified above,
$f_1(x)$ has a Gaussian process prior specified as in Theorem
\ref{thm1}, and $\mbf{b}$ follows a normal distribution with mean
zero and variance-covariance matrix $\mbf{B}$, we can derive that
the joint distribution of $\sbf{y}$ and
$f_0(x)+f_1(x)+\mbf{z}^{T}\mbf{b}$ follows a Gaussian distribution
with mean 0 and covariance matrix
\begin{equation}
\begin{pmatrix}
b\mbf{F}\mbf{V}^{+}\mbf{F}^{T}+\tau^{2}\mbf{S}\mbf{S}^{T}+\sigma^{2}\mbf{I}
& b\mbf{F}\mbf{V}^{+}\tilde{\sbf{\xi}}+\tau^{2}\mbf{S}\sbf{\phi}
\\
b{\tilde{\sbf{\xi}}}^{T}\mbf{V}^{+}\mbf{F}^{T}+\tau^{2}\sbf{\phi}^{T}\mbf{S}^{T}
&
b{\tilde{\sbf{\xi}}}^{T}\mbf{V}^{+}\tilde{\sbf{\xi}}+\tau^{2}\sbf{\phi}^{T}\sbf{\phi}
\end{pmatrix}
\label{cov}
\end{equation}
where $\tilde{\sbf{\xi}}=(R_1 (s_1,x),\dots,R_1
(s_T,x),\mbf{z})^{T}$ is $(T+p) \times 1$, $\sbf{\phi}$ is $m \times
1 $ with the $\nu$th entry $\phi_{\nu}(t)$,
$\mbf{F}=(\mbf{R},\mbf{Z})$, and $V^{+}$ is the Moore-Penrose
inverse of $\mbf{V}=\text{diag}(\mbf{Q},\frac{1}{N{\lambda}
}\mbf{\Omega})$ satisfying
$\mbf{V}\mbf{V}^{+}\mbf{F}^{T}=\mbf{F}^{T}$.

Standard calculation yields
\begin{align*}
E[\mu(x)+\mbf{z}^{T}\mbf{b}|\sbf{y}]
&=(b\tilde{\sbf{\xi}}^{T}\mbf{V}^{+}\mbf{F}^{T}+\tau^{2}\sbf{\phi}^{T}\mbf{S}^{T})
(b\mbf{F}\mbf{V}^{+}\mbf{F}^{T}+\tau^{2}SS^{T}+\sigma^{2}\mbf{I})^{-1}\sbf{y} \\
&=\rho\sbf{\phi}^{T}\mbf{S}^{T}(\mbf{W}+\rho{\mbf{S}}\mbf{S}^{T})^{-1}\sbf{y}
+\tilde{\sbf{\xi}}^{T}\mbf{V}^{+}\mbf{F}^{T}(\mbf{W}+\rho{\mbf{S}}\mbf{S}^{T})^{-1}\sbf{y},
\end{align*}
where $\rho=\tau^{2}/b$, $N{\lambda} =\sigma^{2}/b$, and
$\mbf{W}=\mbf{F}\mbf{V}^{+}\mbf{F}^{T}+N{\lambda} {\mbf{I}}$. Now
letting $\rho\rightarrow\infty$, we have
\begin{align}
\lim_{\rho\rightarrow\infty}&(\rho{\mbf{S}}\mbf{S}^{T}+\mbf{W})^{-1}
=\mbf{W}^{-1}-\mbf{W}^{-1}\mbf{S}(\mbf{S}^{T}\mbf{W}^{-1}\mbf{S})^{-1}\mbf{S}^{T}\mbf{W}^{-1},\label{diff1:2}\\
\lim_{\rho\rightarrow\infty}&\rho{\mbf{S}}^{T}(\rho{\mbf{S}}\mbf{S}^{T}+\mbf{W})^{-1}
=(\mbf{S}^{T}\mbf{W}^{-1}\mbf{S})^{-1}\mbf{S}^{T}\mbf{W}^{-1}.
\end{align}
See \citeN{wahba:83} and \citeN{gu:02} for the proof.

 Therefore,
$\lim_{\tau^{2}\rightarrow\infty}E[\mu(x)+\mbf{z}^{T}\mbf{b}|\sbf{y}]
=\sbf{\phi}^{T}\mbf{d}+\tilde{\sbf{\xi}}^{T}\tilde{\mbf{c}}$,  where
\begin{equation}\label{cd}
\mbf{d}=(\mbf{S}^{T}\mbf{W}^{-1}\mbf{S})^{-1}\mbf{S}^{T}\mbf{W}^{-1}\sbf{y},
\tilde{\mbf{c}}=\mbf{V}^{+}\mbf{F}^{T}(\mbf{W}^{-1}-\mbf{W}^{-1}\mbf{S}(\mbf{S}^{T}\mbf{W}^{-1}\mbf{S})^{-1}\mbf{S}^{T}\mbf{W}^{-1})\sbf{y}.
\end{equation}

It is straightforward to verify that the $\mbf{d}$ and
$\tilde{\mbf{c}}$ given in (\ref{cd}) satisfy (\ref{norm}).

\noindent {\it Proof of Theorem \ref{thm2} :}

The posterior variance can be easily calculated by using expression
(\ref{cov}) as follows,
\[\text{var}[\mu(x)+\mbf{z}^{T}\mbf{b}|\sbf{y}]=\tilde{\sbf{\xi}}^{T}\mbf{V}^{+}\tilde{\sbf{\xi}}+\rho\sbf{\phi}^{T}\sbf{\phi}-(\tilde{\sbf{\xi}}^{T}\mbf{V}^{+}\mbf{F}^{T}+\rho\sbf{\phi}^{T}\mbf{S}^{T})(\mbf{W}+\rho{\mbf{S}}\mbf{S}^{T})^{-1}(\mbf{F}\mbf{V}^{+}\tilde{\sbf{\xi}}+\rho{\mbf{S}}\sbf{\phi})
\]

Notice that
$\lim_{\rho\rightarrow\infty}\rho{I}-\rho^{2}\mbf{S}^{T}(\rho{\mbf{S}}\mbf{S}^{T}+\mbf{W})^{-1}\mbf{S}
=(\mbf{S}^{T}\mbf{W}^{-1}\mbf{S})^{-1}$, and
$\mbf{V}\mbf{V}^{+}\mbf{F}^{T}=\mbf{F}^{T}$. Therefore as
$\rho\rightarrow\infty$, we have
\begin{multline}\notag
\lim_{\tau^{2}\rightarrow\infty}\text{Var}[\mu(x)+\mbf{z}^{T}\mbf{b}|\sbf{y}]/b
=\tilde{\sbf{\xi}}^{T}\mbf{V}^{+}\tilde{\sbf{\xi}}+\sbf{\phi}^{T}(\mbf{S}^{T}\mbf{W}^{-1}\mbf{S})^{-1}\sbf{\phi}-2\sbf{\phi}^{T}(\mbf{S}^{T}\mbf{W}^{-1}\mbf{S})^{-1}\mbf{S}^{T}\mbf{W}^{-1}\mbf{F}\mbf{V}^{+}\tilde{\sbf{\xi}}\notag\\
\quad-\tilde{\sbf{\xi}}^{T}\mbf{V}^{+}\mbf{F}^{T}(\mbf{W}^{-1}-\mbf{W}^{-1}\mbf{S}(\mbf{S}^{T}\mbf{W}^{-1}\mbf{S})^{-1}\mbf{S}^{T}\mbf{W}^{-1})\mbf{F}\mbf{V}^{+}\tilde{\sbf{\xi}}.\notag
\end{multline}

\bibliographystyle{chicago}

\bibliography{root}

\begin{thebibliography}{}

\bibitem[\protect\citeauthoryear{Arbeitman, Furlong, Imam, Johnson, Null,
  Baker, Krasnow, Scott, Davis, and White}{Arbeitman
  et~al.}{2002}]{arbeitman:02}
Arbeitman, M., E.~Furlong, F.~Imam, E.~Johnson, B.~H. Null, B.~S. Baker,
  M.~Krasnow, M.~P. Scott, R.~W. Davis, and K.~P. White (2002).
\newblock Gene expression during the life cycle of drosophila melanogaster.
\newblock {\em Science\/}~{\em 297\/}(5590), 2270--2275.

\bibitem[\protect\citeauthoryear{Castillo-Davis and Hartl}{Castillo-Davis and
  Hartl}{2003}]{cristian:03}
Castillo-Davis, C. and D.~Hartl (2003).
\newblock Genemerge: post-genomic analysis, data-mining and hypothesis.
\newblock {\em Bioinformatics\/}~{\em 19}, 891--892.

\bibitem[\protect\citeauthoryear{Craven and Wahba}{Craven and
  Wahba}{1979}]{craven:79}
Craven, P. and G.~Wahba (1979).
\newblock Smoothing noisy data with spline functions: Estimating the correct
  degree of smoothing by the method of generalized cross-validation.
\newblock {\em Numer. Math.\/}~{\em 31}, 377--403.

\bibitem[\protect\citeauthoryear{Dempster, Laird, and Rubin}{Dempster
  et~al.}{1977}]{em:77}
Dempster, A.~P., N.~M. Laird, and D.~B. Rubin (1977).
\newblock Maximum likelihood from incomplete data via the {EM} algorithm.
\newblock {\em J. Roy. Statist. Soc. Ser. B\/}~{\em 39}, 1--37 (with
  discussions).

\bibitem[\protect\citeauthoryear{Dennis and Schnabel}{Dennis and
  Schnabel}{1996}]{dennis:96}
Dennis, J.~E. and R.~B. Schnabel (1996).
\newblock {\em Numerical Methods for Unconstrained Optimization and Nonlinear
  Equations}.
\newblock Philadelphia: SIAM.
\newblock Corrected reprint of the 1983 original.

\bibitem[\protect\citeauthoryear{Fraley and Raftery}{Fraley and
  Raftery}{1990}]{fraley:2002}
Fraley, C. and A.~E. Raftery (1990).
\newblock Model-based clustering, discriminant analysis, and density
  estimation.
\newblock {\em J. Amer. Statist. Assoc.\/}~{\em 97}, 611--631.

\bibitem[\protect\citeauthoryear{Green}{Green}{1990}]{Green:1990}
Green, P.~J. (1990).
\newblock On the use of the {EM} algorithm for penalized likelihood estimation.
\newblock {\em J. Roy. Statist. Soc. Ser. B\/}~{\em 52}, 443--452.

\bibitem[\protect\citeauthoryear{Gu}{Gu}{2002}]{gu:02}
Gu, C. (2002).
\newblock {\em Smoothing Spline {ANOVA} Models}.
\newblock New York: Springer-Verlag.

\bibitem[\protect\citeauthoryear{Gu and Ma}{Gu and Ma}{2005}]{gm:02}
Gu, C. and P.~Ma (2005).
\newblock Optimal smoothing in nonparametric mixed effect models.
\newblock {\em Ann. Statist.\/}~{\em 33}, 1357--1379.

\bibitem[\protect\citeauthoryear{Hastie and Tibshirani}{Hastie and
  Tibshirani}{1990}]{ht:90}
Hastie, T. and R.~Tibshirani (1990).
\newblock {\em Generalized Additive Models}.
\newblock London: Chapman \& Hall.

\bibitem[\protect\citeauthoryear{Heard, Holmes, and Stephens}{Heard
  et~al.}{2006}]{heard:06}
Heard, N.~A., C.~C. Holmes, and D.~A. Stephens (2006).
\newblock A quantitative study of gene regulation involved in the immune
  response of anopheline mosquitoes: An application of bayesian hierarchical
  clustering of curves.
\newblock {\em J. Amer. Statist. Assoc.\/}~{\em 101\/}(473), 18--29.

\bibitem[\protect\citeauthoryear{Hubert and Arabie}{Hubert and
  Arabie}{1985}]{huber:1985}
Hubert, L. and P.~Arabie (1985).
\newblock Comparing partitions.
\newblock {\em J. Classification\/}~{\em 2}, 193--218.

\bibitem[\protect\citeauthoryear{James and Sugar}{James and
  Sugar}{2003}]{Jame:Suga:2003}
James, G.~M. and C.~A. Sugar (2003).
\newblock Clustering for sparsely sampled functional data.
\newblock {\em J. Amer. Statist. Assoc.\/}~{\em 98\/}(462), 397--408.

\bibitem[\protect\citeauthoryear{Jiang, Ryu, Kiraly, Duke, Reinke, and
  Kim}{Jiang et~al.}{2001}]{jiang:01}
Jiang, M., J.~Ryu, M.~Kiraly, K.~Duke, V.~Reinke, and S.~K. Kim (2001).
\newblock Genome-wide analysis of developmental and sex-regulated gene
  expression profiles in caenorhabditis elegans.
\newblock {\em Proc. Natl. Acad. Sci.\/}~{\em 98\/}(1), 218--223.

\bibitem[\protect\citeauthoryear{Kim and Gu}{Kim and Gu}{2004}]{kg:02}
Kim, Y.-J. and C.~Gu (2004).
\newblock Smoothing spline gaussian regression: More scalable computation via
  efficient approximation.
\newblock {\em J. Roy. Statist. Soc. Ser. B\/}~{\em 66}, 337--356.

\bibitem[\protect\citeauthoryear{Lai, Kosorukoff, Burke, and Kwast}{Lai
  et~al.}{2006}]{lai:2006}
Lai, L.~C., A.~L. Kosorukoff, P.~Burke, and K.~E. Kwast (2006).
\newblock Metabolic-state-dependent remodeling of the transcriptome in response
  to anoxia and subsequent reoxygenation in saccharomyces cerevisiae.
\newblock {\em Eukaryot Cell\/}~{\em 5}, 1468--89.

\bibitem[\protect\citeauthoryear{Liu, Chen, and Wong}{Liu
  et~al.}{1998}]{Liu:Chen:Wong:1998}
Liu, J.~S., R.~Chen, and W.~H. Wong (1998).
\newblock Rejection control and sequential importance sampling.
\newblock {\em J. Amer. Statist. Assoc.\/}~{\em 93}, 1022--1031.

\bibitem[\protect\citeauthoryear{Luan and Li}{Luan and Li}{2003}]{luan:03}
Luan, Y. and H.~Li (2003).
\newblock Clustering of time-course gene expression data using a mixed-effects
  models with {B}-spline.
\newblock {\em Bioinformatics\/}~{\em 19\/}(4), 474--282.

\bibitem[\protect\citeauthoryear{Luan and Li}{Luan and Li}{2004}]{luan:04}
Luan, Y. and H.~Li (2004).
\newblock Model-based methods for identifying periodically regulated genes
  based on the time course microarray gene expression data.
\newblock {\em Bioinformatics\/}~{\em 20\/}(4), 332--339.

\bibitem[\protect\citeauthoryear{Ma, Wang, Tenorio, de~Hoop, and van~der
  Hilst}{Ma et~al.}{2007}]{Ma:2007}
Ma, P., P.~Wang, L.~Tenorio, M.~V. de~Hoop, and R.~D. van~der Hilst (2007).
\newblock Imaging of structure at and near the core mantle boundary using a
  generalized {R}adon transform: 2. statistical inference of singularities.
\newblock {\em J. Geophys. Res.\/}~{\em 112}, B08303.

\bibitem[\protect\citeauthoryear{McCarroll, Murphy, Zou, Pletcher, Chin, Jan,
  Kenyon, Bargmann, and Li}{McCarroll et~al.}{2004}]{haoli:04}
McCarroll, S.~A., C.~T. Murphy, S.~Zou, S.~D. Pletcher, C.~S. Chin, Y.~N. Jan,
  C.~Kenyon, C.~I. Bargmann, and H.~Li (2004).
\newblock Comparing genomic expression patterns across species identifies
  shared transcriptional program in aging.
\newblock {\em Nature Genetics\/}~{\em 36\/}(2), 197--204.

\bibitem[\protect\citeauthoryear{McLachlan and Peel}{McLachlan and
  Peel}{2001}]{McLa:Peel:fini:2001}
McLachlan, G.~J. and D.~Peel (2001).
\newblock {\em Finite Mixture Models}.
\newblock John Wiley \& Sons.

\bibitem[\protect\citeauthoryear{Nychka}{Nychka}{1988}]{nychka:88}
Nychka, D. (1988).
\newblock Bayesian confidence intervals for smoothing splines.
\newblock {\em J. Amer. Statist. Assoc.\/}~{\em 83}, 1134--1143.

\bibitem[\protect\citeauthoryear{Ramsay and Silverman}{Ramsay and
  Silverman}{2002}]{Rams:Silv:appl:2002}
Ramsay, J.~O. and B.~W. Silverman (2002).
\newblock {\em Applied Functional Data Analysis: Methods and Case Studies}.
\newblock Springer-Verlag Inc.

\bibitem[\protect\citeauthoryear{Ramsay and Silverman}{Ramsay and
  Silverman}{2005}]{Rams:Silv:func:2005}
Ramsay, J.~O. and B.~W. Silverman (2005).
\newblock {\em Functional Data Analysis}.
\newblock Springer-Verlag Inc.

\bibitem[\protect\citeauthoryear{Robinson}{Robinson}{1991}]{robin:91}
Robinson, G.~K. (1991).
\newblock That {BLUP} is a good thing: The estimation of the random effects.
\newblock {\em Statist. Sci.\/}~{\em 6}, 15--51 (with discussions).

\bibitem[\protect\citeauthoryear{Robinson, Grigull, Mohammad, and
  Hughes}{Robinson et~al.}{2002}]{funspec:2002}
Robinson, M.~D., J.~Grigull, N.~Mohammad, and T.~R. Hughes (2002).
\newblock Funspec: a web-based cluster interpreter for yeast.
\newblock {\em BMC Bioinformatics\/}~{\em 3}, 3--35.

\bibitem[\protect\citeauthoryear{Silverman and Wood}{Silverman and
  Wood}{1987}]{silv:wood:1987}
Silverman, B.~W. and J.~T. Wood (1987).
\newblock The nonparametric estimation of branching curves.
\newblock {\em J. Amer. Statist. Assoc.\/}~{\em 82}, 551--558.

\bibitem[\protect\citeauthoryear{Spellman, Sherlock, Zhang, Iyer, Anders,
  Eisen, Brown, and Botstein D~Futcher}{Spellman et~al.}{1998}]{spellman:1998}
Spellman, P.~T., G.~Sherlock, M.~Q. Zhang, V.~R. Iyer, K.~Anders, M.~B. Eisen,
  P.~O. Brown, and B.~Botstein D~Futcher (1998).
\newblock Comprehensive identification of cell cycle-regulated genes of the
  yeast saccharomyces cerevisiae by microarray hybridization.
\newblock {\em Mol Biol Cell.\/}~{\em 9\/}(12), 3273--97.

\bibitem[\protect\citeauthoryear{Storey, Xiao, Leek, and Tompkins}{Storey
  et~al.}{2005}]{storey:05}
Storey, J.~D., W.~Xiao, J.~T. Leek, and R.~Tompkins, R. G.and~Davis (2005).
\newblock Significance of time course microarray experiments.
\newblock {\em Proc. Natl. Acad. Sci.\/}~{\em 102}, 12837--12842.

\bibitem[\protect\citeauthoryear{Wahba}{Wahba}{1983}]{wahba:83}
Wahba, G. (1983).
\newblock Bayesian ``confidence intervals'' for the cross-validated smoothing
  spline.
\newblock {\em J. Roy. Statist. Soc. Ser. B\/}~{\em 45}, 133--150.

\bibitem[\protect\citeauthoryear{Wahba}{Wahba}{1990}]{wahba:90}
Wahba, G. (1990).
\newblock {\em Spline Models for Observational Data}, Volume~59 of {\em \rm
  {CBMS-NSF} Regional Conference Series in Applied Mathematics}.
\newblock Philadelphia: SIAM.

\bibitem[\protect\citeauthoryear{Wang, de~Hoop, van~der Hilst, Ma, and
  Tenorio}{Wang et~al.}{2006}]{Wang:2006}
Wang, P., M.~V. de~Hoop, R.~D. van~der Hilst, P.~Ma, and L.~Tenorio (2006).
\newblock Imaging of structure at and near the core mantle boundary using a
  generalized radon transform: 1. construction of image gathers.
\newblock {\em J. Geophys. Res.\/}~{\em 111}, B12304.

\bibitem[\protect\citeauthoryear{Wang}{Wang}{1998}]{wang:98a}
Wang, Y. (1998).
\newblock Mixed-effects smoothing spline {ANOVA}.
\newblock {\em J. Roy. Statist. Soc. Ser. B\/}~{\em 60}, 159--174.

\bibitem[\protect\citeauthoryear{Wei and Tanner}{Wei and
  Tanner}{1990}]{Wei:Tann:1990}
Wei, G.~C. and M.~A. Tanner (1990).
\newblock A {M}onte {C}arlo implementation of the {EM} algorithm and the poor
  man's data augmentation algorithms.
\newblock {\em J. Amer. Statist. Assoc.\/}~{\em 85}, 699--704.

\bibitem[\protect\citeauthoryear{Zhang, Lin, Raz, and Sowers}{Zhang
  et~al.}{1998}]{Zhan:Lin:Raz:Sowe:semi:1998}
Zhang, D., X.~Lin, J.~Raz, and M.~Sowers (1998).
\newblock Semiparametric stochastic mixed models for longitudinal data.
\newblock {\em J. Amer. Statist. Assoc.\/}~{\em 93}, 710--719.

\end{thebibliography}

\end{document}